\documentclass{ifacconf}

\usepackage{graphicx}      % include this line if your document contains figures
\usepackage{natbib}        % required for bibliography
\usepackage{pgf}
\usepackage{tabularx}
\usepackage{epstopdf}
\usepackage{epsfig}
\usepackage{graphics} % for pdf, bitmapped graphics files
\usepackage{epstopdf}
\usepackage{enumerate}
\usepackage{caption}
\usepackage{amsmath} % assumes amsmath package installed
\usepackage{amssymb, dsfont,mathabx}  % assumes amsmath package installed
\usepackage{subfig,pgfplots}
\usepackage{multirow}
\usepackage{xcolor,colortbl}

\newtheorem{remark}{Remark}%}
\graphicspath{{images/}}

\newcommand{\R}{\mathbb{R}}

\newcommand{\N}{\mathbb{N}}

\newcommand{\mc}[1]{\mathcal{#1}}
\newcommand{\mf}[1]{\mathfrak{#1}}

\DeclareMathOperator*{\argmin}{argmin}

\newcommand{\col}{\mathrm{col}}
\newcommand{\tup}[1]{\textup{#1}}

\usepackage{mathtools}
\usepackage[acronym]{glossaries}
\newacronym{MIP}{MIP}{Mixed-Integer Programming}
\newacronym{SoC}{SoC}{State of Charge}
\newacronym{PEV}{PEV}{Plug-in Electric Vehicle}
\newacronym{EV}{EV}{Electric Vehicle}
\newacronym{GS}{GS}{Gauss-Southwell}
\newacronym{MINE}{$\varepsilon$-MINE}{$\varepsilon$-Mixed-Integer Nash Equilibrium}
\newacronym{CTM}{CTM}{Cell Transmission Model}
\newacronym{CTMs}{CTM-\textit{s}}{ Cell Transmission Model with service station}
\newacronym{CS}{CS}{Charging Station}
\newacronym{FV}{FV}{fuel vehicle}
\newacronym{r2s}{$\mathrm{r2s}$}{road-to-station}
\newacronym{s2r}{$\mathrm{s2r}$}{station-to-road}
\newacronym{HO}{HO}{Highway Operator}
\newacronym{TDM}{TDM}{Traffic Demand Management}
\newacronym{ATDM}{ATDM}{Active Traffic Demand Management}
\newacronym{NDW}{NDW}{Nationaal
Dataportaal Wegverkeer}
\newacronym{TTT}{TTT}{Total Travel Time}
\newacronym{NLP}{NLP}{Nonlinear Programming}
\newacronym{MINLP}{MI-NLP}{Mixed Integer Nonlinear Programming}
\newacronym{ST}{ST}{Service Station}
\newacronym{NN}{NN}{Neural Network}
\newacronym{GA}{GA}{Genetic Algorithm}
\newacronym{MLSE}{MLSE}{Mean Logarithmic Squared  Error}

\usepackage{bbold}
\usepackage{balance}

\begin{document}
\begin{frontmatter}

\title{Optimal service station design for traffic mitigation via genetic algorithm  and neural network\thanksref{footnoteinfo}} 

\thanks[footnoteinfo]{The work of Cenedese  and Lygeros was supported by NCCR Automation, a National Centre of
Competence in Research, funded by the Swiss National Science
Foundation (grant number $180545$).}

\author[ETH]{Carlo Cenedese} 
\author[UNIPV]{Michele Cucuzzella}
\author[UNIPV]{Adriano Cotta Ramusino}
\author[UNIPV]{Davide Spalenza}
\author[ETH]{John Lygeros}
\author[UNIPV]{Antonella Ferrara}

\address[ETH]{Department of Information Technology and Electrical Engineering, ETH Z\"urich, Zurich, Switzerland ({\texttt{ccenedese@ethz.ch}}).}
\address[UNIPV]{Department of Electrical, Computer and Biomedical Engineering, University of Pavia, Pavia, Italy ({\texttt{michele.cucuzzella@unipv.it}}).}

\begin{abstract}                % Abstract of not more than 250 words.
This paper analyzes how the presence of service stations on highways affects traffic congestion. We focus on the problem of optimally designing a service station to achieve beneficial effects in terms of total traffic congestion and peak traffic reduction.
Microsimulators cannot be used for this task due to their computational inefficiency. We propose a genetic algorithm based on the recently proposed \gls{CTMs}, that efficiently describes the dynamics of a service station. Then, we leverage the algorithm to train a neural network capable of solving the same problem, avoiding implementing the \gls{CTMs}.  
Finally, we examine two case studies to validate the capabilities and performance of our algorithms. In these simulations, we use real data extracted from Dutch highways.
 \end{abstract}

\end{frontmatter}
%===============================================================================

\section{Introduction}
During the last  decades, traffic congestion has been worsening in major cities around the world creating  costs not only in terms of waiting for resources but also in terms of emissions.
According to the INRIX $2021$ Global Traffic Scorecard report, the U.S. drivers wasted on average $99$ hours a year due to traffic congestion, generating  nearly $\$88$ billion in total costs. The situation was even worse in  France, where in $2021$ drivers in Paris spent on average 140 hours in traffic~\cite{inrix2021global}. 

To manage rush hour traffic, a wide variety of tools have been proposed in the literature as well as implemented in practice. Arguably, one of the most  effective approaches is the optimal design  of the traffic infrastructure with the goal of mitigating the traffic congestion.
On highways, an important factor influencing the flow of vehicles is the presence of \glspl{ST}. Initially, authors focused on the modelling and optimally control of \gls{ST} by employing queuing theory, see \citep{wang:1995:service_station_model}. More recently, the spread of electric vehicles renew the interest of the research community towards service stations, and in particular those offering  the possibility to recharge \gls{PEV}. 
\cite{ferro:2020:bilevel_EV_charging} propose a bi-level where the higher level aims at computing the optimal planning of charging stations. The lower level instead is modelled as  a traffic assignment problem among commuters.
The problem of finding the optimal location of charging stations has been also studied in \citep{kong:2017:EV_optimal_location,KONG:2019:optimal_location_and_planning}, where the authors consider a simple traffic model interconnected with the power grid to determine where the charging station should be located.  Refer to \citep{pagany:2019:spatial_loc_of_CS} for a survey on spatial localization methodologies for the \gls{PEV} charging infrastructure.

In \citep{gusrialdi:2017:scheduling_EV}, the authors propose a control scheme to mitigate queues at charging stations optimizing the drivers experience. An incentive based control is presented in  
\citep{cenedese:2022:bottleneck_incentive}, where a dynamic discount on the charging of \gls{PEV} is able to decrease traffic congestion over a bottleneck. 
Finally, in \citep{cenedese:2020:highway_control_pI}, the authors incentivize electric vehicles to stop at a \gls{ST} during periods of peak congestion via a discount on the purchased energy. This is an online control of a \gls{ST} base on game theory (similarly to \citep{cenedese:2019:PEV_MIG}) and it is able to decrease the peak of traffic congestion.

The majority of the contributions in the literature focuses more on charging stations rather than the more general \glspl{ST}. Moreover, they do not consider the effect that such stations have on the mainstream traffic dynamics and the possibility that a careful design can mitigate traffic congestion. In this work, we bridge this gap formulating the problem of optimal design of a \gls{ST} with the objective of minimizing the overall traffic congestion. The main contribution of the paper can be summarized as follows:
\begin{itemize}
\item we formalize the problem of the optimal design of a \gls{ST} to mitigate traffic congestion using the \gls{CTMs} as a \gls{MINLP};
\item we design a \gls{GA}  able to  compute the optimal solution of the \gls{MINLP} for a  given  highway stretch;
\item we use the \gls{GA} to create a training dataset for different highways configurations and use it to train a \gls{NN}. Once  trained, the \gls{NN} is able to solve the problem  without requiring the user to implement the \gls{CTMs};
\item we analyze two case studies based on  real data and validate the efficacy of the proposed solutions. 
\end{itemize}

%\subsection{Paper organization}
%Paper organization

%===============================================================================

\section{Problem formulation}
\begin{figure}[t]
\centering
\includegraphics[trim=0 170 70 230,clip,width=\columnwidth]{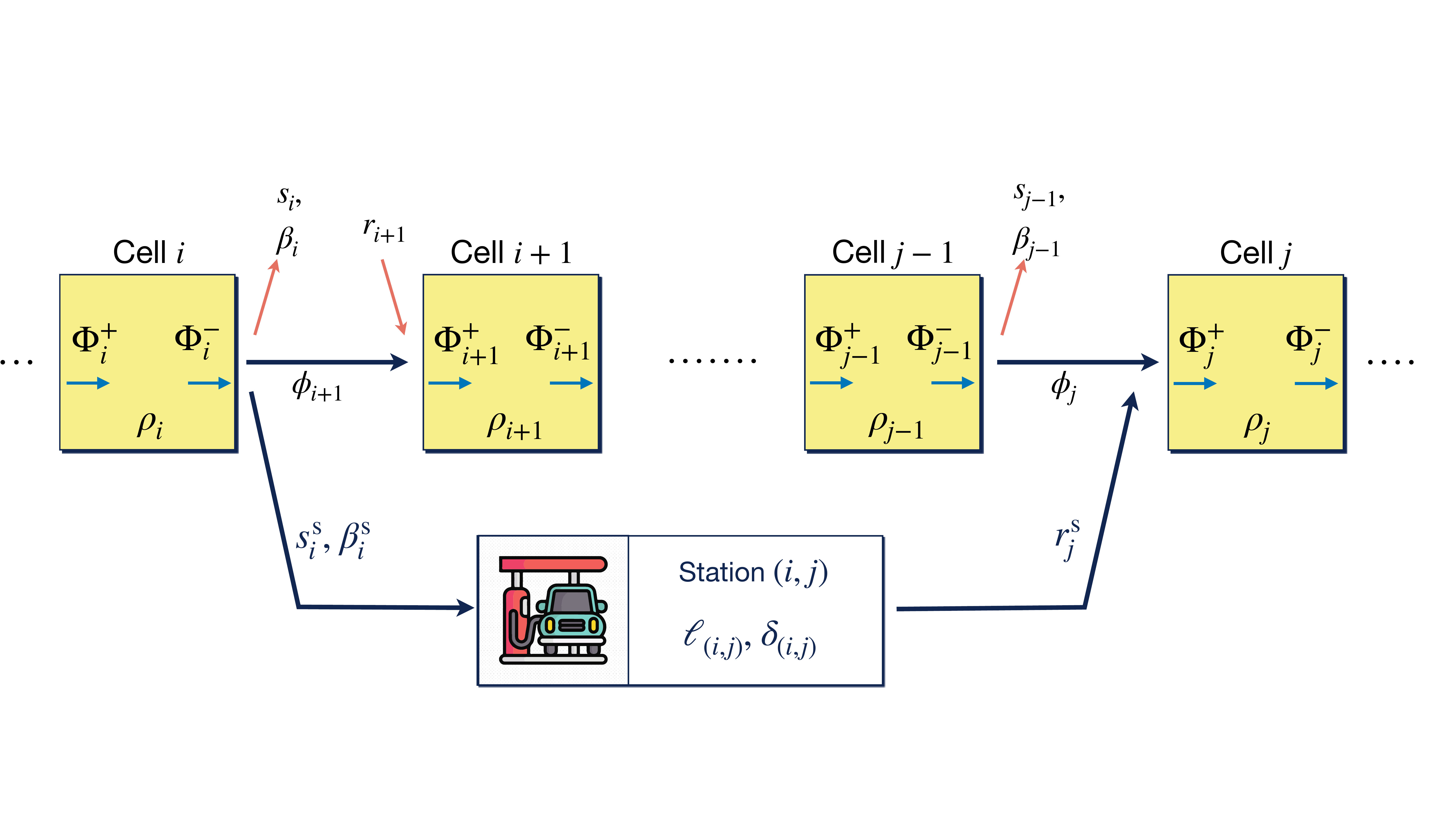}
\caption{The model and variables of the \gls{CTMs}, including on- and off-ramps together with a single \gls{ST} $(i,j)$. }\label{fig:CTMs_scheme}
\end{figure}
In this section, we first propose a brief introduction on the \gls{CTMs} and discuss the  variables involved in the optimal design of a service station. Then, we formalize the problem as a general \gls{MINLP}.

\subsection{\gls{CTMs} overview}
The \gls{CTMs} was recently proposed by~\cite{cenedese:2022:CTMs} as an extension of the classical~\gls{CTM} (see e.g. \citep{daganzo1992CTM2}), where a highway stretch is divided into $N$  consecutive cells described by the index set $\mc N\coloneqq \{1,\cdots,N\}$.   If a driver can enter a \gls{ST}  from cell $i$, called \textit{access cell}, and exit it by merging back into the mainstream in cell $j$, called \textit{exit cell}, then the \gls{ST} is denoted by the ordered couple $p=(i,j)\in \mc M\subseteq \mc N\times \mc N$,  see Figure~\ref{fig:CTMs_scheme}. The cardinality of the set describes the number of service stations, i.e., $M=|\mc M|$.  In this work, we focus on the optimal design of a single unimodal \gls{ST} on a highway stretch, i.e.,  $M=1$ and the drivers at the \gls{ST} are assumed to be homogeneous. This standing assumptions simplify some of the dynamics introduced in \cite[Sec.~II.A]{cenedese:2022:CTMs}. Relaxing these assumptions would not create technical difficulties but rather increase the computational complexity of the problem, and thus it is left for future works.

The intrinsic characteristics of the highway stretch are described by the set of parameters  in Table~\ref{tb:CTMs_param}. To facilitate the discussion, we distinguish the parameters in three different groups according to the type: those associated with the cells (denoted by $P$), those describing the \gls{ST} (denoted by $S$), and finally some other parameter that is assumed to be fixed (denoted by $F$). More precisely, we define the first two types by the following vectors
$$P = \col((L_i,\overline v_i,q_i^{\max},\rho_i^{\max},\beta_i)_{i\in\mc N})\in\R^{6N}$$
and
$$ S = \col(i,j,\delta_{(i,j)},\beta^{\tup s}_i)\in \mc S =\mc N^2\times\R\times [0,1]. $$
The values of the parameters in $F$ are the result of a  reasonable design choice on  the \gls{ST}.
% \red{\begin{remark}
% Add also the discussion on the fact that $S$ can be diminished and put into $F$. Useful later for the mixed scenarios
% \end{remark}}

\begin{table}[t]
\begin{center}
\captionsetup{width=\columnwidth}
\caption{ \gls{CTMs} parameters associated with cell $i\in\mc N $ and \gls{ST} $p=(i,j)$ divided by type: $P$, $S$, or $F$.}\label{tb:CTMs_param}
\begin{tabular}{c|ccl}
 & Name & Unit & Description \\\hline
\multirow{6}{*}{$P$}&$L_i$ & [\tup{km}] & cell length  \\
&$\overline v_i$  & [\tup{km/h}] & \text{free-flow speed}  \\
&$w_i$ & $[\tup{km/h}]$ & congestion wave speed\\
&$q_i^{\max}$ & $[\tup{veh/h}]$ & maximum cell capacity \\
&$\rho_i^{\max}$&$[\tup{veh/km}]$& maximum jam density \\
&$\beta_i$ & \% &   off-ramp split ratio\\
\hline
\multirow{4}{*}{$S$}&$i$ &  &  \gls{ST}'s access cell \\
&$j$ &  &  \gls{ST}'s exit cell\\
&$\delta_{(i,j)}$ & [h] &  avg. time spent at the \gls{ST}\\
&$\beta^{\tup s}_i$ & \% & \gls{ST}'s off-ramp split ratio\\\hline
\multirow{3}{*}{$F$}&$p_i^{\tup{ms}}$ & &   priority of mainstream $i$\\
&$r^{\tup s,\max}_{j}$ & $[\tup{veh/h}]$ & \gls{ST}'s maximum on-ramp capacity \\
&$L_{(i,j)}$&  & \gls{ST}'s length in number of cells \\\hline
\end{tabular}
\end{center}
\end{table}

The traffic flow dynamics evolve over $T_{\tup h}$ time intervals $k$ of length $T$ and indexed by the set $\mc T\coloneqq \{1,\cdots,T_{\tup h}\}$. The mainstream and  \gls{ST} dynamics are described via the \gls{CTMs} variables listed in Table~\ref{tb:CTMs_var}. 
The density $\rho_i$ of cell $i$  is described by
\smallskip
\begin{equation}
\label{eq:rho_dyn}
\rho_i(k+1) = \rho_i(k) + \dfrac{T}{L_i}\left( \Phi_i^+(k)-\Phi_i^-(k)\right) \:,
\end{equation}
where the inflow and outflow are defined respectively as
\smallskip
\begin{subequations}
\label{eq:Phi_+_-}
\begin{align}
\label{eq:Phi_-}
\Phi^-_i(k) &=
 \phi_{i+1}(k) + s_i(k) + s^{\tup s}_i(k)\\ 
\Phi^+_i(k) &= \phi_i(k) + r_i(k) +  r^{\tup s}_i(k).
\end{align} 
\end{subequations}
Note that throughout the paper we consider  $\phi_1(k)$ and the demand of the on-ramps  as  known external inputs to the dynamical system.  

The flow of vehicles entering the \gls{ST} and exiting the highway via an off-ramp from cell $i$ are simply defined via the associated splitting ratios, i.e., $s^{\tup s}_{i}(k)= \beta^{\tup s}_{i}\Phi^-_{i}(k)$ and $s_{i}(k)= \beta_{i}\Phi^-_{i}(k)$ where $\beta^{\tup s}_{i}+\beta_{i}\leq 1$. 
\begin{table}[t]
\captionsetup{width=\columnwidth}
\caption{ \gls{CTMs} variables for the dynamics at time $k\in\N$ associated  with cell $i\in\mc N$ and \gls{ST}  $p=(i,j)$.}\label{tb:CTMs_var}
\begin{center}
\begin{tabular}{ccl}
Name & Unit & Description \\\hline
$\rho_i(k)$ & $[\textup{veh/km}]$ & traffic density of  cell $i$\\
$\Phi_i^{+}(k) (\Phi_i^{-}(k)) $&$[\textup{veh/h}]$ &  total flow entering (exiting) cell $i$\\
$\phi_i(k)$&$[\textup{veh/h}]$ &  flow entering cell $i$ from  $i-1$\\
$r_i(k)$&$[\textup{veh/h}]$ & flow merging into $i$ via an on-ramp\\
$s_i(k)$&$[\textup{veh/h}]$ & flow leaving $i$ via an off-ramp\\\hline
$\ell_{(i,j)}(k)$ & $\:[\textup{veh}]$ &  vehicles currently at the \gls{ST}\\
$e_{(i,j)}(k)$ & $\:[\textup{veh}]$ &  vehicles queuing  to exit the \gls{ST}\\
$s^{\tup s}_i(k)$&$[\textup{veh/h}]$& flow leaving  $i$ to enter the \gls{ST}\\
$r^{\tup s}_j(k)$&$[\textup{veh/h}]$ &  flow  merging into  $j$ from the \gls{ST}\\
\hline
\end{tabular}
\end{center}
\end{table}
The dynamics of the \gls{ST} $p=(i,j)$ are described by: the number of vehicles at the \gls{ST} during the time interval $k$, i.e., 
\begin{align}
\label{eq:ell}
    \ell_{p}(k+1)=\ell_{p}(k)+T\left[ s_{i}^{\tup s}(k) - r^{\tup s}_{j}(k) \right],
\end{align}
and among the vehicles at the \gls{ST}, by the ones waiting for merging back into the mainstream, i.e.,
\begin{equation}\label{eq:e}
     e_p(k+1) = e_p(k) + T\left[ s_{i}^{\tup s}(k-\delta_{p}) - r^{\tup s}_{j}(k)\right]\,.
 \end{equation}
The number of vehicles that  attempt to exit the \gls{ST} during $k$ is $s^{\tup s}_{i}(k-\delta_{p})+\frac{e_p(k)}{T}$, while the demand of vehicles that try to merge back into the mainstream reads as
\begin{equation}
\label{eq:demad_ij}
    D^{\tup{s}}_{p}(k) = \min\left(s^{\tup s}_{i}(k-\delta_{p})+\frac{e_p(k)}{T},r_{j}^{\tup s,\max} \right) .
\end{equation}

Due to space limitations, we omit the rest of the dynamics derivation that can be found in \cite[Eq.~7-14]{cenedese:2022:CTMs}, i.e, the exact definitions of the demand and supply among cells in the case of free-flow and congested scenarios that in turn are used to compute $\phi_i(k)$ and $r^{\tup s}_i(k)$, respectively. 

By denoting all the \gls{CTMs} variables in Table~\ref{tb:CTMs_var} as $x$, we define the dynamics of the whole model in compact form as 
\begin{equation}\label{eq:compact_CTMs_dyn}
x(k+1) = f(x(k),S,P,F),
\end{equation}
where $f$ is a highly nonlinear parametric function that incorporate the dynamics in \eqref{eq:rho_dyn}--\eqref{eq:demad_ij} and \cite[Eq.~7-14]{cenedese:2022:CTMs}. The dependency from $\phi_1$ and the demand of the on-ramps have been omitted to ease the notation.

\subsection{General \gls{MINLP} formulation}\label{ssec:NLP_formulation}

In this section, we formalize the problem of optimal \gls{ST} design and cast it as a \gls{MINLP}.

\subsubsection{Cost function:}
We are now ready to elaborate the concept of optimality considered in this work. As previously stated, the effect that a \gls{ST} has on the traffic conditions is relevant and thus its design can alleviate or exacerbate the traffic congestion on the highway stretch.  To quantify such effect, we introduce two cost components, i.e., $\xi_\Delta$ and $\pi_\Delta$, which are both related to the additional travel time due to the traffic congestion on the highway stretch during $k\in\mc T$, i.e., 
\begin{equation}
    \label{eq:Delta}
    \Delta(k) \coloneqq \sum_{i\in\mc N} \left(\dfrac{L_i}{v_i(k)} - \dfrac{L_i}{\overline v_i} \right) \in\R_+.
\end{equation}  
Notice that  $\Delta(k)=0$ implies that the highway is operating in free-flow conditions. 

The first cost component corresponds to the area under $\Delta$, hence
\begin{equation}\label{eq:xi_Delta}
\xi_\Delta(S|P) = T \sum_{k\in\mc T} \Delta(k)\in \R_+,
\end{equation}
 and it captures an aggregate information on the duration and intensity of the traffic congestion. From \eqref{eq:Delta}, it follows that if $\xi_\Delta=0$, then there is no congestion during $\mc T$. This quantity is a non-normalized version of the Relative Congestion Index (RCI) used in~\citep{afrin:2020:RCI_survey,tnag:2018:RCI}.

% \red{\begin{remark}[Weighted $\xi_\Delta$]
% Say that is can be weighted with $L_i\rho_i(k)$ such that it is more important to minimize $\xi_\Delta$ when a lot of people are affected by it.
% \end{remark}}
On top of minimizing $\xi_\Delta$, we also aim at achieving a peak-shaving effect on the traffic congestion. For this reason the second cost component captures the percentage of peak congestion reduction due to the (beneficial) presence of the  \gls{ST}, and it is defined as follows 
\begin{equation}
    \label{eq:pi_Delta}
    \pi_\Delta(S|P) = \dfrac{\max_k(\Delta_0(k)) - \max_k(\Delta(k))}{\max_k(\Delta_0(k))}\leq 1,
\end{equation}
where $\Delta_0$ is computed as in \eqref{eq:Delta}, considering no \gls{ST} on the highway stretch. Here, $\pi_\Delta=1$ implies that there is a complete elimination of the peak, i.e., the highway operates in free-flow conditions. On the other hand, if $\pi_\Delta\leq 0$ there is no beneficial effect in introducing the \gls{ST} in terms of peak reduction.  

Then, the final cost function used in the optimization problem reads as 
\begin{equation}\label{eq:cost_NLP}
    c(S|P) = \alpha \xi_\Delta(S|P)-\pi_\Delta(S|P),
\end{equation}
where $\alpha$ is a normalizing coefficient. Specifically, if $\alpha= 1/(\sum_i{\textstyle \frac{L_i}{\overline v_i}})$, then the first component of $c$ corresponds to  the RCI. 

\subsubsection{Constraints:}
The choice of the feasible \gls{ST}'s parameters $S$ is limited by a set of constraints that are necessary to ensure that the attained results have a clear physical interpretation. In the remainder, given a vector $y$, we denote its $q$-th component by $[y]_q$. 

First, we ensure that the length  of the  \gls{ST}  is  exactly equal to $L_{(i,j)}$. Then, the parameters $S\in\mc S$ must satisfy
\begin{equation}
    \label{eq:cnst_len}
    [S]_2-[S]_1 = L_{(i,j)}.
\end{equation}
%Notice that this condition does not imply a \gls{ST} that has always the exact physical length since the cells are heterogeneous in $L_i$.
The rest are box constraints limiting the values of $S$. We can describe in compact form all the above constraints  as $AS\leq b$, for a suitable choice of $A$ and $b$.

\begin{remark}
Given the particular configuration of the  highways or preferences, different constraints can be added. For example, the access point of the \gls{ST} can be constrained to be only in those cells that do not already have a on- or off-ramp.  
\end{remark}

We are now ready for   formulating the problem of designing the optimal \gls{ST} for a highway stretch associated with a specific $P$,  i.e., computing $S^\star$  by solving the following \gls{MINLP} problem
\begin{align}
    \label{eq:NLP}
    &S^\star  =& &\argmin_{S\in\mc S} &&c(S|P) \\\nonumber
    &        && \quad\tup{s.t.} &&AS\leq b\\\nonumber
    &        && &&x(k+1) = f(x(k),S,P,F),\:\forall k\in\mc T \\\nonumber
    &        && && x(0) = x_0,
\end{align}
where $x_0$ are the dynamics' initial conditions. The presence of integer variables in $S$ (the first two components) and the highly nonlinar dynamics of the \gls{CTMs} with respect to $S$  make the problem complex from both a theoretical and computational point of view.  Although  approximations that  simplify the classical \gls{CTM} dynamics have been proposed in \citep{zilia:2000:LP_CTM,hong:2001:CTM_control}, these modifications do not simplify much  problem \eqref{eq:NLP} since the optimization is performed with respect to $S$.

% \section{Solution methods}
% \label{sec:sol_methods}
\section{Genetic Algorithm}
\label{ssec:ga}
The first solution we  present is based on the \gls{GA} that has been successfully applied in the past also by other authors working on traffic control to attain numerical solutions to \lq\lq untractable'' optimization problems,  see \citep{hong:2001:dyn_net}.

The structure of the \gls{GA} is simple and very effective. First, a set of candidate maximizers of a given fitness function are selected, then their fitness is computed, and finally a new ``generation'' of candidates is obtained by creating variations of  the best performing ones, refer to \citep{Katoch:2021:GA_review} for an a detailed  overview on the topic.    

The first population $\mathfrak S_1$ is composed of $N_{\tup{GA}}$ candidates randomly chosen in the set of feasible design parameters, that is 
\begin{equation}
    \label{eq:feasible_set}
    \left\{ S\in\mc S\,|\, AS\leq b \right\}. 
\end{equation} 
The fitness function is simply the opposite of the cost in \eqref{eq:cost_NLP}, hence, for a given $(P,F)$ and $x_0$, the fitness of $S\in\mathfrak{S}_i$ in generation $i$ is $-c(S|P)$. 

Given a generation $j$ and the set of $N_{\tup{GA}}^\star\leq N_{\tup{GA}}$ best performing elements $\mf S_j^\star\subseteq \mf S_j$,  the new population $\mf S_{j+1}$ includes the $N_{\tup{GA}}^\star$ best performing elements of  generation $j$, i.e., $\mf S_j^\star$, and also their offspring and mutations. 
 The offspring exploits the best performers of the previous generation, i.e., $\mf S_j^\star$ and are generated via double-point crossover. 
On the other hand, the  mutations, chosen randomly, make the \gls{GA} explore the space of feasible parameters in \eqref{eq:feasible_set}, reducing the probability of converging to local minima. In $\mf S_{j+1}$, $N_{\tup{GA}} - N_{\tup{GA}}^\star$ elements are composed by the offspring and mutations, and among these the probability of mutation is denoted by $p_{\tup{GA}}$.   Note that in general we cannot guarantee that the new elements   satisfy the constraints. Then, this issue can be overcome in two ways: the unfeasible solutions are replaced by new feasible ones, or a very low fit can be assigned to unfeasible solutions such that the algorithm is forced to not select them.  
The \gls{GA} stops when the number of contiguous generations with no improvement of the fitness exceeds $K_{\tup{stop}}$. % It is important to notice that there is no guarantee that all the offspring and mutations of $\mf S_j^\star$ satisfy \eqref{eq:feasible_set}. 

%  \red{there are different kind of mutations that can be applied which one is the one selected? And what is the formula associated like the re is some sort of coefficient of mutation?}

% \blue{
% %The solution we chose is instead a genetic algorithm, that iteratively searches the best solution to the problem, and reaches the optimum 1-2 minutes. Starting from an initial population of inputs $S$, either randomly generated or set by the user, the algorithm obtains the respective outputs via the model under exam, for us the CTM-s, and evaluates them based on the fitness function. The fittest among the inputs are taken to be parents of a new generation, by combining their features. The fitness function has been taken equal to $-C(S \mid P, F)$.\\
% The algorithm can be altered with some optional steps that can steer the execution more towards speed or stability. Our choices in this respect will be explained in section \ref{sec:sims}, as well as the checks performed to ensure that the solution returned by the GA is indeed optimal.
% }

\subsection{Implementation}
We validate the discussed \gls{GA}  by comparing it with a ``brute force'' algorithm that  solves \eqref{eq:NLP} by sampling the set of feasible $S$ and find the optimal solution by inspection. Due to the combinatorial nature of the brute force algorithm, we consider highway stretches with a  moderate number of cells and we perform a moderate number of comparison between the two algorithms.  Namely, we use 5 different choices of $P$ for $N\in\{10,15,20\}$ attaining a total number of 15 different values of $P$ used to validate the \gls{GA}. 
As expected,  the brute force algorithm computational time grows rapidly with $N$,  see Table~\ref{tb:exec_time}.
\begin{table}[b]
\captionsetup{width=\columnwidth}
\begin{center}
\caption{Execution time}\label{tb:exec_time}
\begin{tabular}{c|ccc}
N &$10$ & $15$ & $20$  \\\hline
avg. time brute force & $1.39$ [h] & $2.26$ [h]  & $3.14$ [h]  \\
avg. time \gls{GA} & $0.13$ [h] & $0.31$ [h]  & $0.42$ [h]  \\\hline
\end{tabular}
\end{center}
\end{table}

We denote this validation set by $\mf P$; the parameters of each cell composing $P\in\mf P$ are randomly selected within the intervals in Table~\ref{tb:P_values}. The selected values are in line  with those identified in the case study presented in Section~\ref{sec:case_study}. The value of $\alpha$ has been chosen equal to $0.01$.
\begin{figure}[t]
\centering
\includegraphics[width=\columnwidth]{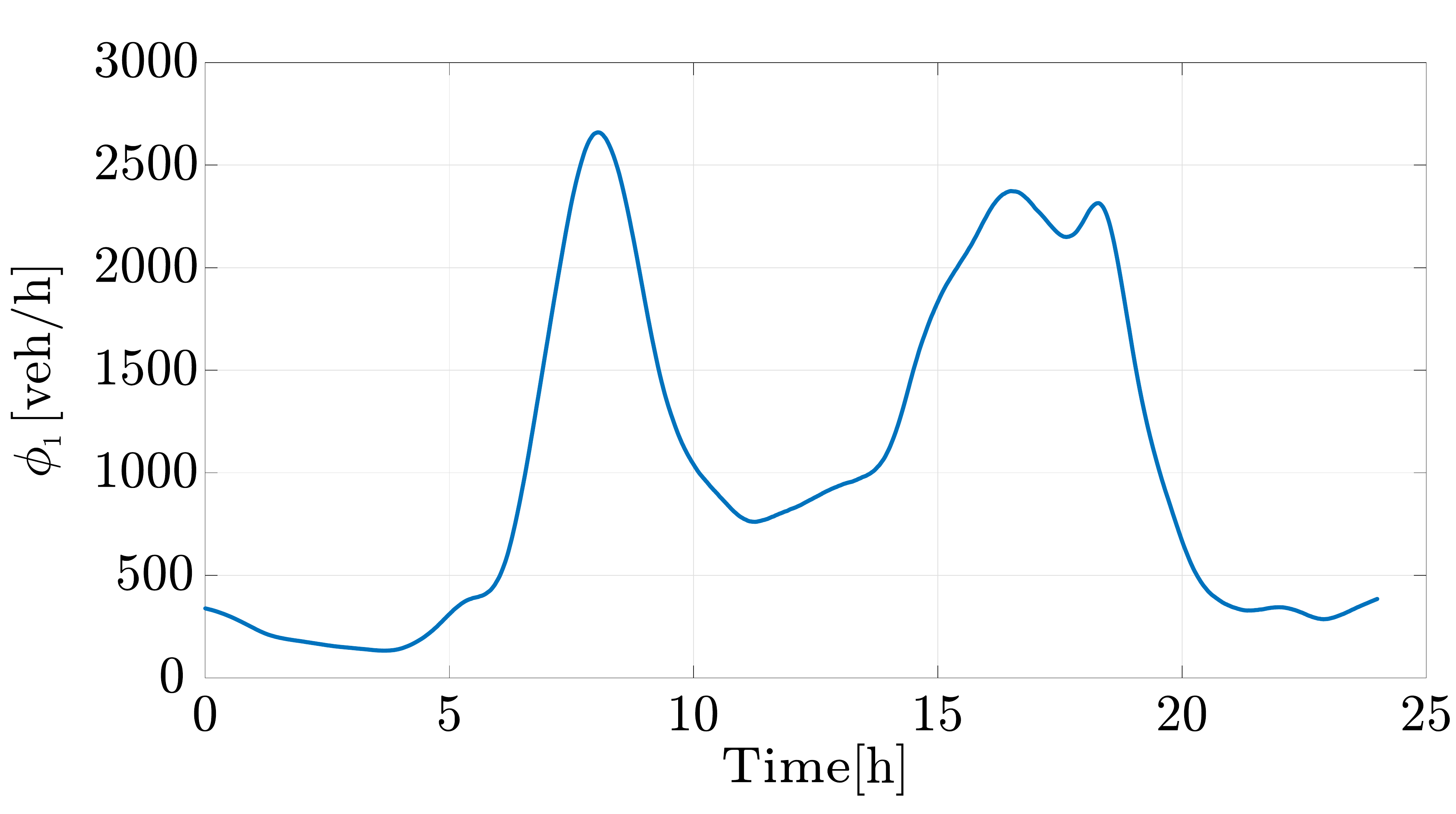}
\caption{Input $\phi_1(k)$ used for the simulations.}
\label{fig:phi1}
\end{figure}

The value of $\phi_1(k)$ in Figure~\ref{fig:phi1} is used for all the simulations, and it is a smooth version of the flow in a highway line during a typical working day. Here, we considered highway stretches in which there are no on- and off-ramps. As anticipated, the values of $F$ has been considered fixed and equal to 
\begin{equation}\label{eq:F_val}
     F=\col(0.95,\,1500,\,2).
\end{equation}
\begin{table}[b]
\captionsetup{width=\columnwidth}
\begin{center}
\caption{Range of values used to generate the different $P\in\mf P$.}
\resizebox{\columnwidth}{!}{%
\begin{tabular}{ccccc}\label{tb:P_values}
$L_i$[m] &  
$\overline v_i$ [km/h]  &
$w_i$[km/h] & 
$q_i^{\max}$ [veh/h] & 
$\rho_i^{\max}$[veh/km]  \\ \hline
$[300, 1000]$ & 
$[80, 110]$ &
$[10, 40]$   &
$[1500, 2500]$  & 
$[70, 100]$ \\ \hline
\end{tabular}
}
\end{center}
\end{table}

We implemented the \gls{GA} in Python using the PyGAD, the parameters of the algorithm are reported in Table~\ref{tb:GA_param}. The \gls{GA}'s parameters  are kept constant for all $P\in\mf P$.
\begin{table}[h]
\captionsetup{width=\columnwidth}
\begin{center}
\caption{\gls{GA} parameters}\label{tb:GA_param}
\begin{tabular}{cccc}
$N_{\tup{GA}}$ & $N_{\tup{GA}}^\star$ & $p_{\tup{GA}}$ & $K_{\tup{stop}}$ \\\hline
16 & 4 & 0.1 & 7 \\\hline
\end{tabular}
\end{center}
\end{table}
The range of values from which the components of $S\in\mf S_1$ are randomly drawn are reported in Table~\ref{tb:S_values}, the same intervals are used as upper and lower bounds constraints for all $S\in\mf S_j$ and  $j\in \N$.  
\begin{table}[h]
\captionsetup{width=\columnwidth}
\begin{center}
\caption{Range of values for $S\in\mf S_j$ and $j\in\N$.}
\label{tb:margins}
\begin{tabular}{ccc}\label{tb:S_values}
$(i,j)$  & $\delta_{(i,j)}$ [min] & $\beta^{\tup s}_i$ \\\hline
 $\mc N\times \mc N$ & $[0, 720]$
 & $[0, \,0.2]$ \\\hline
\end{tabular}
\end{center}
\end{table}

\begin{figure}[t]
\centering
\includegraphics[width=\columnwidth]{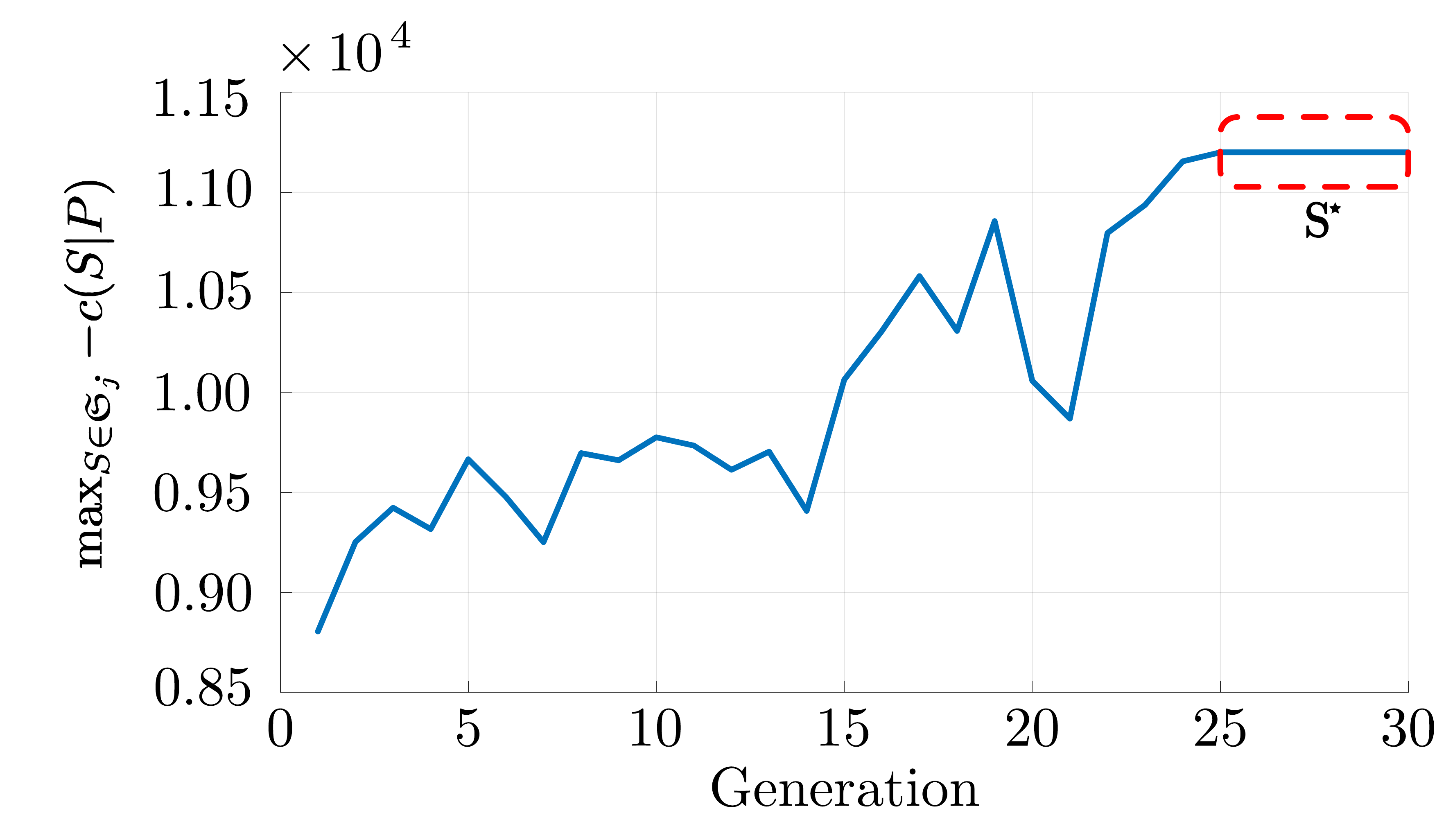}
    \caption{Fitness of the best performing $S\in\mf S_j$ for each generation $j$. The plateau after generation $25$ is where we identify $S^\star$.}\label{fig:fitness_GA}
\end{figure}
\begin{figure}[t]
\centering
\includegraphics[width=\columnwidth]{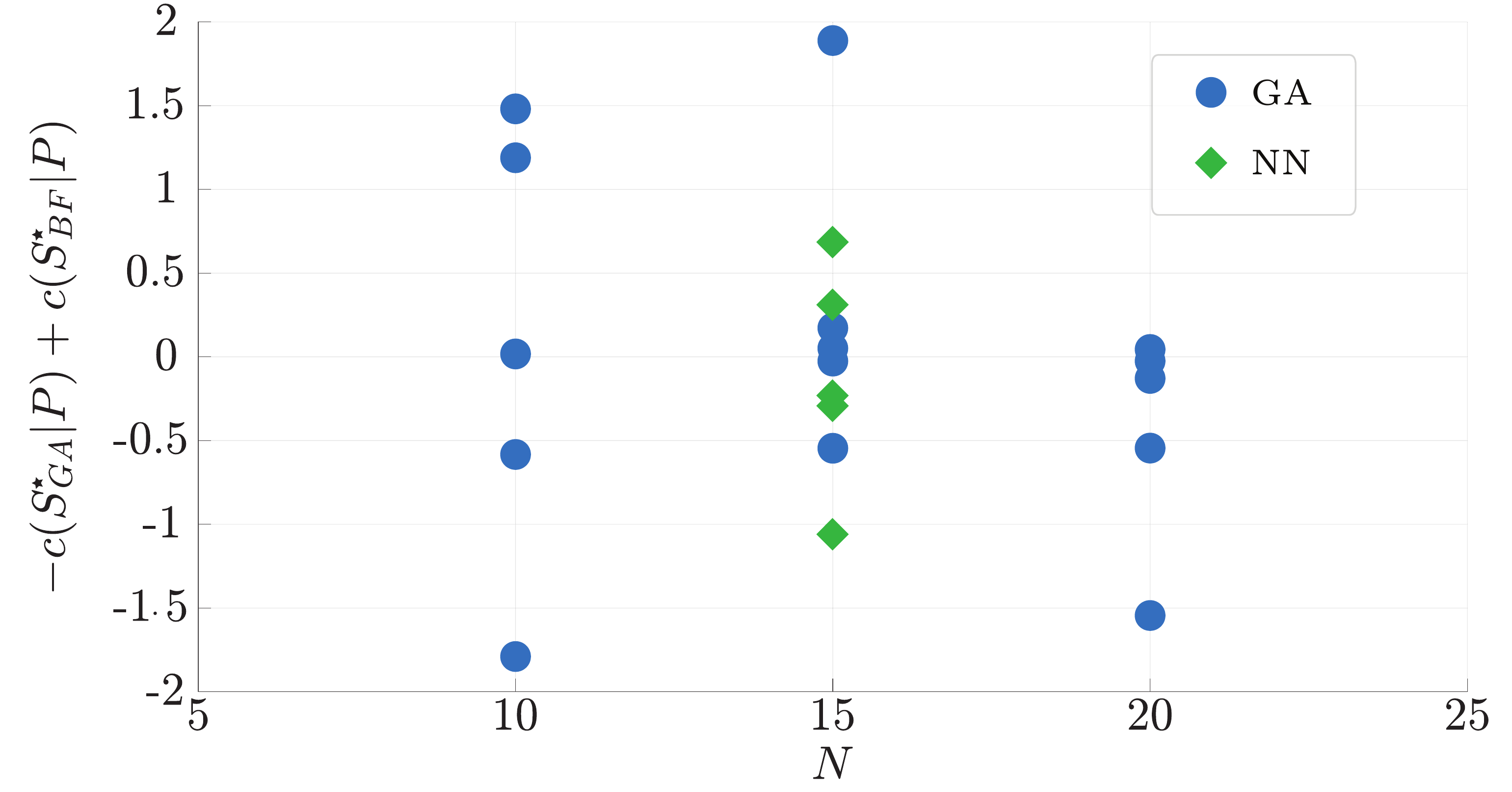}
\caption{The difference between the maximum fitness value computed via the brute force algorithm and the one attained via \gls{GA} (in blue) and via \gls{NN} (in blue) for different values of $P\in\mf P$.} \label{fig:S_GA_NN}
\end{figure}

As shown in Figure~\ref{fig:fitness_GA}, the \gls{GA} takes only $30$ generations to converge to $S^\star_\tup{GA}$. We compare the cost  $S^\star_\tup{GA}$  with the one achieved by the  brute force solution $S^\star_\tup{BF}$. In Figure~\ref{fig:S_GA_NN}, we plot for all $P\in\mf P$ the difference $c(S^\star_\tup{BF}|P)-c(S^\star_\tup{GA}|P)$.  The result obtained by the \gls{GA} is comparable to the one obtained via the brute force algorithm. In fact, the  difference between the two values stays in the interval $[-2,2]$ even though the cost is usually on the order of hundreds. Therefore, the \gls{GA} can successfully compute a solution to \eqref{eq:NLP} that achieves performance similar to those attained via the brute force algorithm. Interestingly, in some circumstances $S^\star_\tup{GA}$ obtains a lower cost than the one obtained by $S^\star_\tup{BF}$. This is due to the sampling on the feasible space of $S$ that in some circumstances is not dense enough.

%  In Figure~\ref{fig:S_GA_NN}, we depict the distance of the optimal solution computed by the \gls{GA} $S_{GA}^\star$ and the one obtained via brute force $S_{BF}^\star$. Notice that to have a more readable comparison we plot $\lVert \overline S_{GA}^\star - \overline S_{BF}^\star\rVert$ where  $\overline S_{GA}^\star$ and $\overline S_{BF}^\star$ are  the normalized versions of the above vectors, respectively. 

% \red{Something on the satisfactory results attained. Maybe also about the attained $\Delta$ or some discussion on how close Aimsun is to these results, not clear yet. We should first see the simulations. } \green{MAgnitudo delle cost functions circa $100$. }
% Finally, the optimality of the solution has been sample checked by running batches of simulations with the CTM-s on large sets of stations that include the optimal one found by the GA and hundreds of variations of it. The solutions found by the GA proved to be consistently the best out of the batches.

\section{Neural Network}
\label{ssec:NN}

One of the drawbacks of the \gls{GA} is its lack of scalability when the length and complexity of the stretches increase. Moreover, it highly depends on the underlying model, i.e. the \gls{CTMs}, and it solves \eqref{eq:NLP} for one specific stretch, i.e., $P$.  In this section, we aim at tackling these shortcomings by deriving an approximation of the mapping between the features of a generic highway stretch ($P$ and $F$), and the associated $S^\star$. We achieve this via a \gls{NN} that requires a single computationally expansive training session but after that it can be used with negligible computational time for generic stretches, refer to  \citep{hagan:1997:NN,abiodub:2018:survey_NN} and references therein.

To train the  \gls{NN}, we need an extensive  dataset of different highways stretches and  optimal \gls{ST}. Real data cannot be used since the location of the \gls{ST} cannot be changed and thus it cannot be assessed whether it is the optimal choice or not.  An alternative can be represented by microsimulators, such as Aimsun Next 22, that are widely regarded as precise representations of real traffic dynamics. However, the execution times are prohibitive. In fact, a simulation of the $24$ hours on a $6$ km highway stretch require $30$ minutes to be completed. Therefore, finding the optimal \gls{ST} design and repeat the process to create dataset of reasonable size is practically unfeasible.

\begin{figure}[t]
\centering
\includegraphics[trim={10 150 90 190},clip,width=\columnwidth]{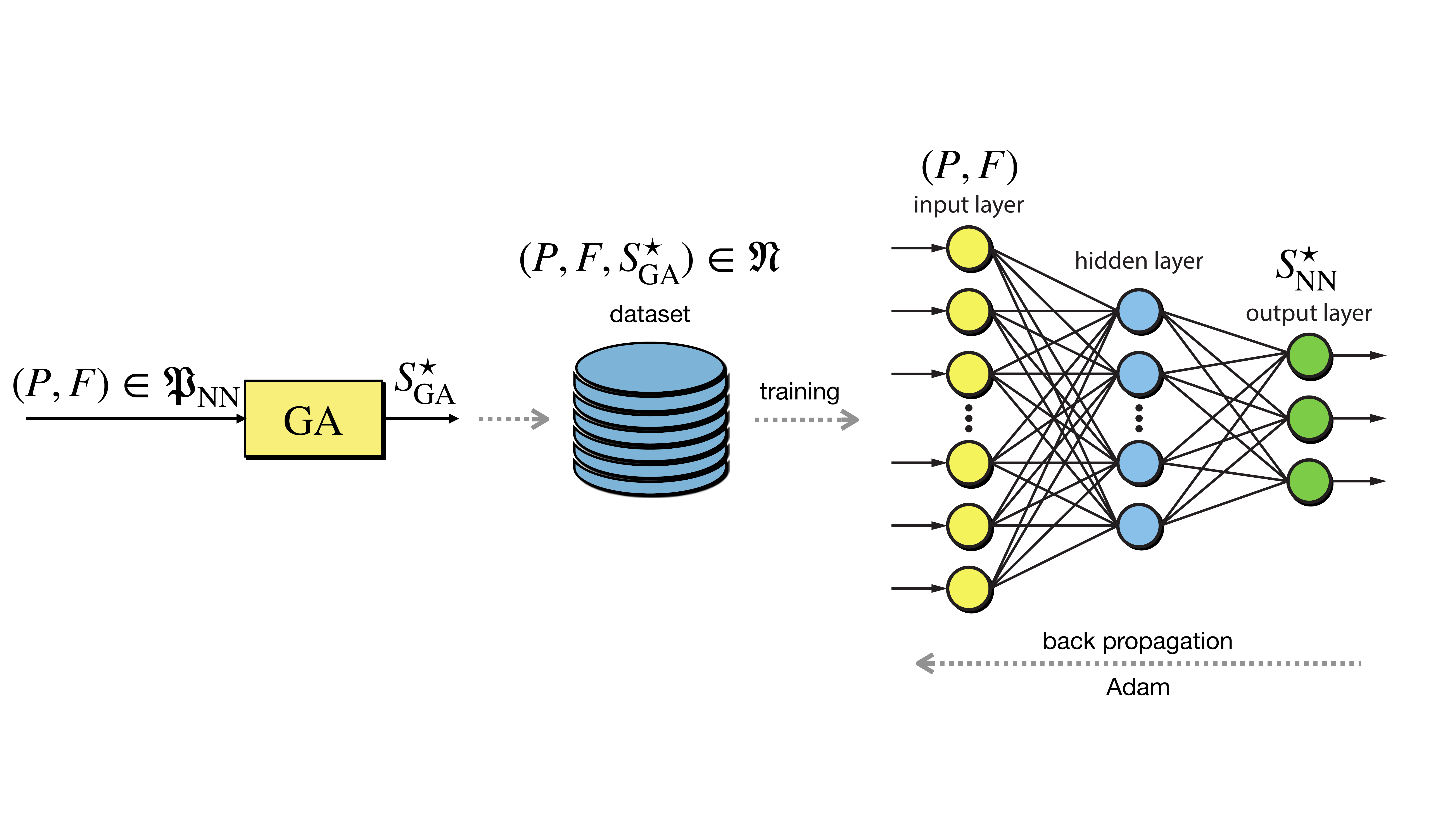}
\caption{The pipeline used to generate the dataset $\mf N$ and train of the \gls{NN} from the initial set $\mf S_{\tup{NN}}$.}\label{fig:NN_training}
\end{figure}

The \gls{CTMs} represents a novel opportunity, which makes it possible to process over $4\cdot 10^4$ days of simulation in less than $24$ hours. Moreover, for each highway stretch, identified by $(P,F)$, we can leverage the \gls{GA} introduced in the previous section to compute   $S^\star_\tup{GA}$, and thus generate a collection of  tuples $(P,F,S^\star_\tup{GA})$ that compose the training dataset $\mf N$, see Figure~\ref{fig:NN_training}. To create a database that comprehends as many scenarios as possible, we vary not only $P$ but also $F$, within reasonable ranges. The set of all these couples $(P,F)$ is denoted by $\mf P_{\tup{NN}}$. The variation of $\phi_1(k)$ and, if present,  of the on-ramps' demand  should be performed carefully. In fact, the choice of  unrealistic demand profiles for a given highway stretch can generate a biased \gls{NN}.  

Our numerical simulations show that a \gls{NN} with a  single hidden layer  is able to learn how to solve \eqref{eq:NLP}, if the dataset $\mf{N}$  captures enough isntances of the \gls{CTMs} dynamics. The input layer is composed of as many layer as the components of  $(P,F)$. Notice that this implies that, for highway stretches with different number of cells $N$, different  \glspl{NN} should be trained. The generalisation of the method to relax this assumption is left for future works.

We select Adam \citep{adam} as backpropagation algorithm. It is an adaptive momentum gradient descent based algorithm able to achieve satisfactory results by  dynamically change the learning rate as the learning progresses. Finally, we use the \gls{MLSE} as  loss function for both the training and validation. 
 
%The $\gls{NN}$ attained after the training is able to approximate the mapping  

% \blue{
% More information on the creation of these stretches can be found in
% \ref{sec:sims}.\\
% The datasets are thus composed as follows:
% \begin{itemize}
%     \item \textbf{inputs}: a large number of random stretches, completely defined by all their CTM-s parameters:\\
%     \begin{align*}
%         P = \{L_i, \overline v_i, w_i, q_i^{\max},\rho_i^{\max}, \green{\beta_i, r_i} \}
%     \end{align*}

%     \item \textbf{outputs}: the features of the optimal station:\\
%     \begin{align*}
%         S^\star = \argmin_S C(P,S), \qquad \forall S \in \mathcal{S}
%     \end{align*}
% \end{itemize}
% \green{This setup allows to include as many or as little parameters as needed, e.g. on and off ramps can be included or omitted and priorities can be optimized or assumed fixed. The implementation shown in section \ref{subsec:nn-impl} disregards ramps and assumes the priorities to be fixed identified values for the sake of simplicity.
% }
% }

\subsection{Implementation}
\label{subsec:nn-impl}
\begin{figure}[h]
\centering
\includegraphics[width=\columnwidth]{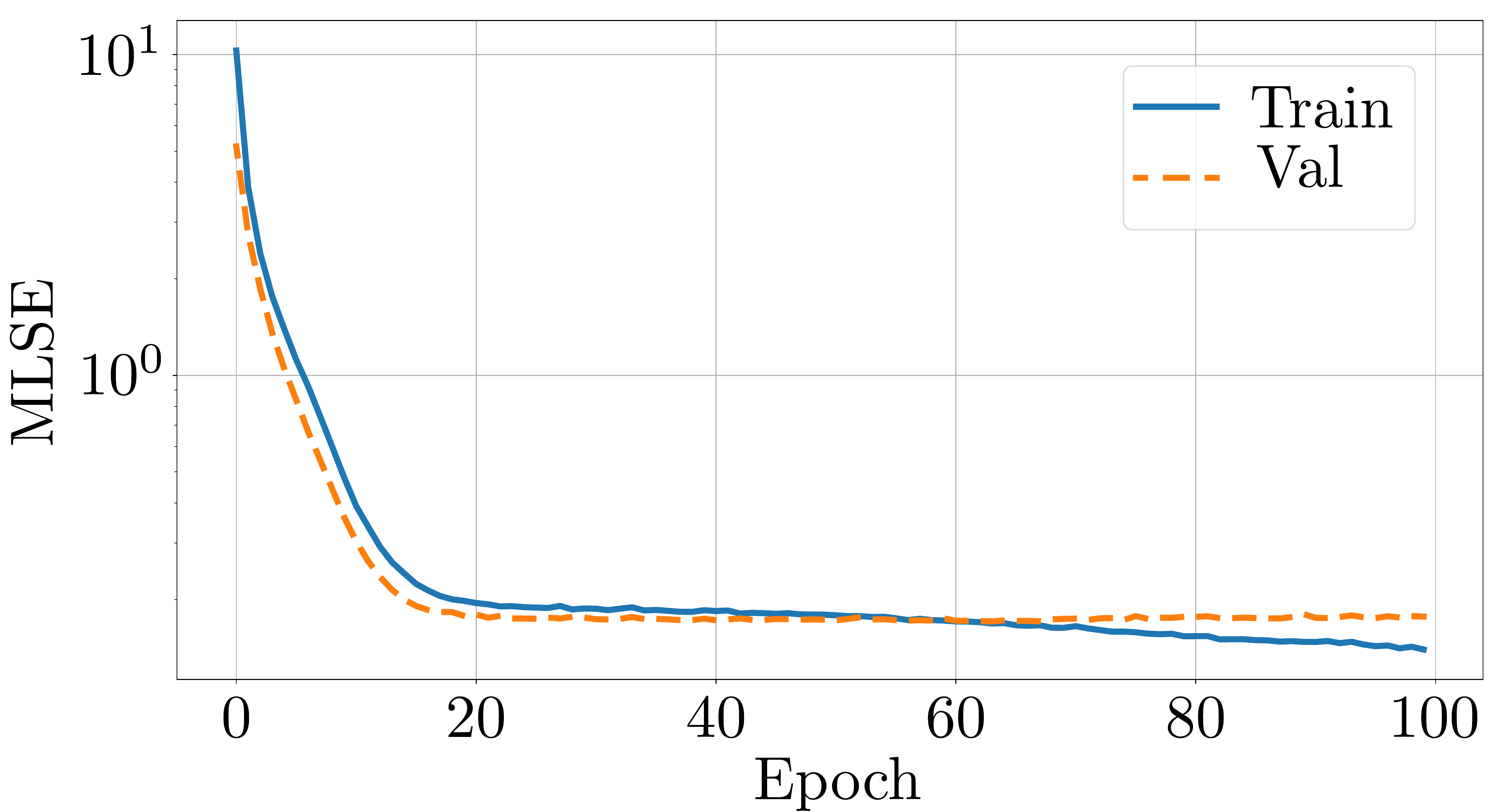}
    \caption{Loss function, i.e., \gls{MLSE} curve, of the \gls{NN} ($1$ hidden layer, $55$ neurons) across $100$ epochs, with batch size of $70$ for both the training and validation phase.}\label{fig:NN_val}
\end{figure}

Next we derive the \gls{NN} as described above; for the implementation we use Tensorflow. % \citep{tensorflow2015-whitepaper}.
To training dataset $\mf N$ is attaiend from a collection of $|\mf S_{\tup{NN}}|=10^4$ highway stretches of $N=15$ cells. Each cell's parameters are randomly drawn from the ranges in  Table~\ref{tab:nn_dataset}.
\begin{table}[h]
\captionsetup{width=\columnwidth}
\begin{center}
    \caption{Range of parameters used to generate $\mf S_{\tup{NN}}$ and then $\mf N$. }
    \label{tab:nn_dataset}
    \begin{tabular}{c|cc}
    \hline
    &{$|\mf S_{\tup{NN}}|$ and $|\mf N|$} & $10^4$ \\ \hline
    \multirow[c]{5}{*}{$P$}& $L_i$ & $[300,\, 1000]$\\ 
    &$\overline v$ & $[80,\,110]$  \\
    &$w$ & $[10,\,40]$  \\
    &$q^{\max} $ & $[1500,\, 2500]$        \\ 
    &$\rho^{\max}$ & $[70,\,100]$           \\ \hline
    \multirow[c|]{3}{*}{$F$} &$p^{\tup{ms}}_i$ & 0.95           \\ 
    &$r^{\tup{s,max}}_j$         & 1500  \\ 
    &$L_{(i,j)}$         & $2$  \\ \hline
\end{tabular}
\end{center}
\end{table}
The \gls{GA} is tuned as in the previous section, i.e., according to Table~\ref{tb:GA_param} and \ref{tb:S_values}.
For the training phase, we tested different batch sizes, i.e., the number of samples that will be propagated through the \gls{NN}. Large batches degrade the \gls{NN}'s ability to generalize while small ones create convergence problems, see Figure~\ref{fig:NN_val}. The best results are obtained for batches of  $[60, 70]$ samples, with an \gls{MLSE} between $0.160$ and $0.162$,  showing that the \gls{NN} is able to compute a solution $S^\star_\tup{NN}$ that is almost identical to the one obtained via the \gls{GA}. This is confirmed  by Figure~\ref{fig:S_GA_NN} where $S^\star_\tup{NN}$, compared to the brute force solution, obtains similar performance to those of the \gls{GA}. %\red{ Say that the MLSE is a result of the algorithm and therefore is good casue small.}\red{DEscribe what is going on in Fig 6 and also refer to Fig 5 for the performance validation.}

\begin{table}[h]
\captionsetup{width=\columnwidth}
\begin{center}
    \caption{ \gls{NN} configuration and \gls{MLSE} values calculate on the validation data set.}
    \label{tab:nn}
    \begin{tabular}{cc}
     \hline
    Hidden layers        & $1$              \\ 
    Neurons hidden layer           & $55$             \\ 
    Dropout hidden layer & $0.2$  \\
    Activation function & ReLU  \\
    Batch size              & $[60 ,\, 70]$        \\ 
    Optimizer           & Adam           \\ 
    Learning rate           & Adaptive           \\ \hline
  
\end{tabular}
\end{center}
\end{table}

\section{Case studies}\label{sec:case_study}

\begin{figure}[h]
\centering
\includegraphics[trim={0 200 0 250},clip,width=\columnwidth]{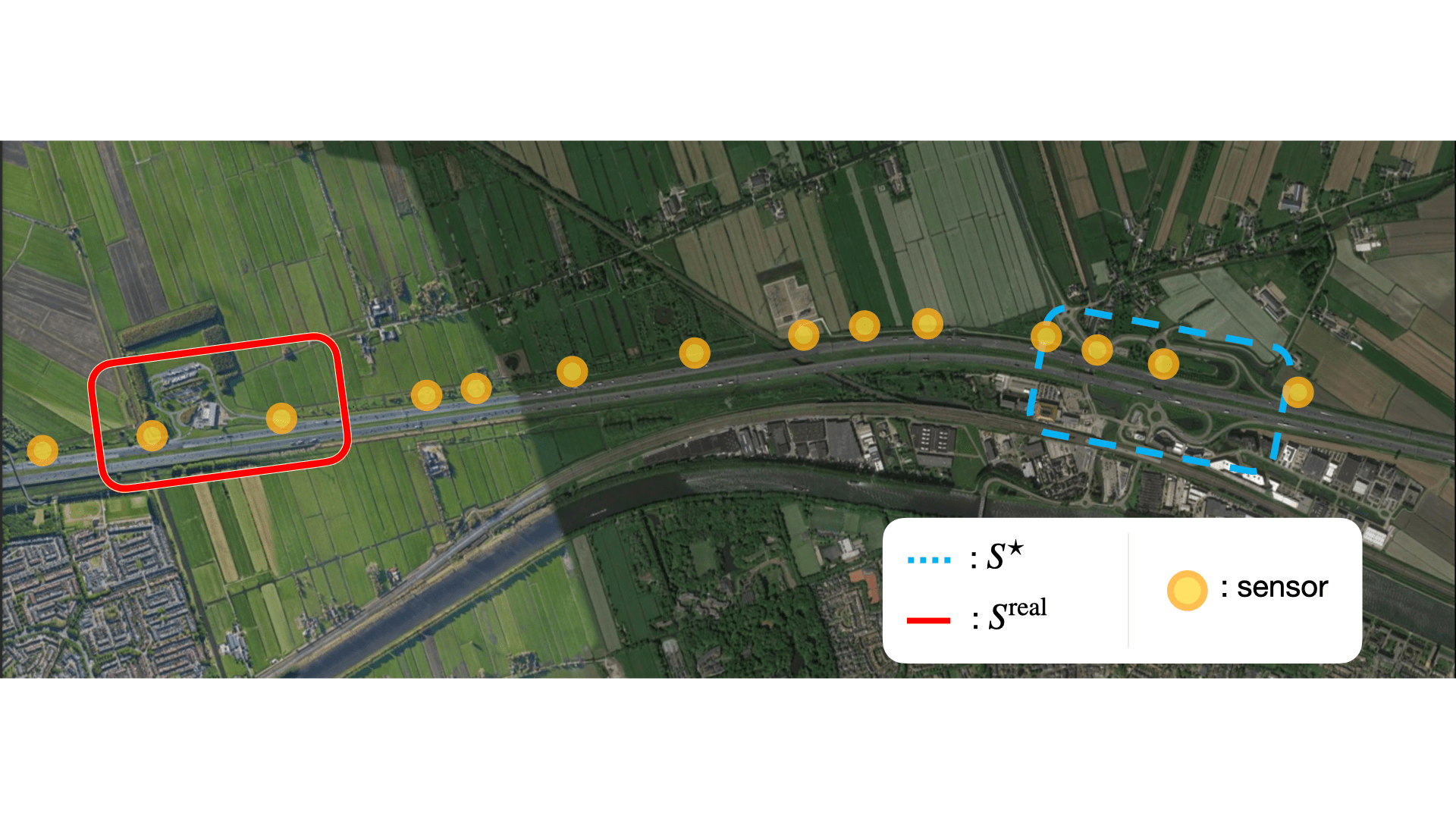}
    \caption{The optimal placement of the A2 \gls{ST} (in dashed blue box), and the actual placement (in the red  box). The sensors considered are highlighted by yellow circles.}\label{fig:A2}
\end{figure}

\begin{figure}[t]
\centering
\includegraphics[trim={0 200 0 250},clip,width=\columnwidth]{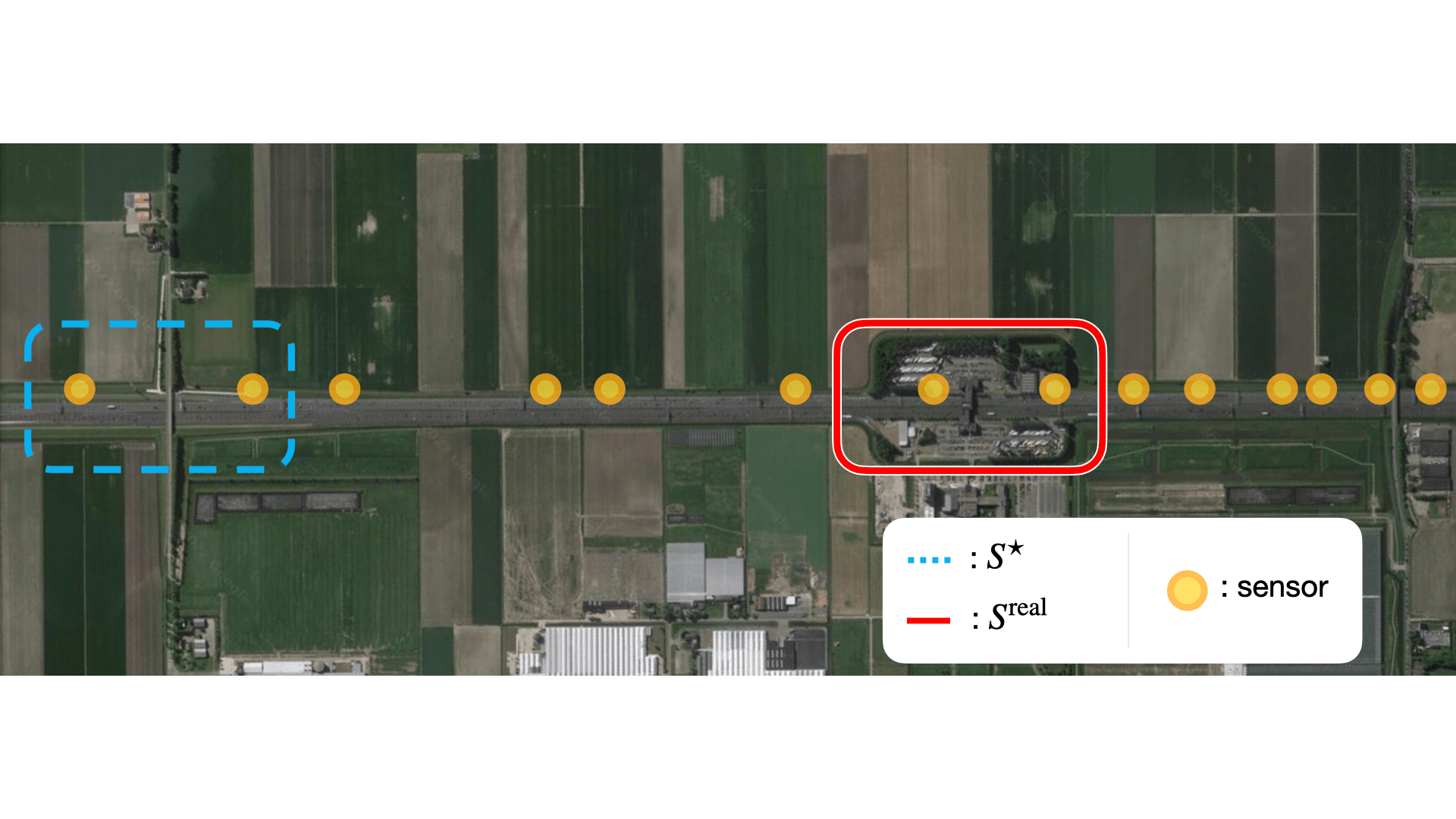}
    \caption{The optimal placement of the A4 \gls{ST} (in dashed blue box), and the actual placement (in the red  box). The sensors  are denoted  by yellow circles.}\label{fig:A4}
\end{figure}

We apply the \gls{GA} and \gls{NN} designed in the previous sections to solve \eqref{eq:NLP} for   two case studies. We compare the optimal \gls{ST}, i.e., $S^\star$, with the one  currently in place $S^{\tup{real}}$ and quantify the benefit that the implementation of the  optimal design  could have on traffic congestion. The real traffic data are extracted from the \gls{NDW} that provide free access to the data associated to all the sensors located on the Dutch highways. The data has been collected for  $24$ hours with a frequency of $1$ minute. 
The highway stretch selected meet the following requirements: at least a \gls{ST} is present and the sensors are placed such  that it is possible to retrieve $s^{\tup s}$ and $r^\tup{s}$.  
The two stretches selected are:
\begin{itemize}
\item $\tup{A}2$ southbound, from Amsterdam towards Utrecht, in Figure~\ref{fig:A2},
\item $\tup{A}4$ eastbound, from Amsterdam towards Den Haag, in Figure~\ref{fig:A4}.
\end{itemize}
Notice that in both cases we did not model the presence of on- and off-ramps, since from our micro-simulations they did not affect the optimal position of the \gls{ST}.
The identification of the cells' parameters, i.e., of $P_{\tup{A2}}$ and $P_{\tup{A4}}$ respectively, has been performed according to~\cite{dervisoglu:2009:aut_CTM_id}, and they are reported in Tables~\ref{tab:a2_cells} and \ref{tab:a4_cells}. The identification of the \gls{ST} parameters, viz. $(F_\tup{A2},S^\tup{real}_\tup{A2})$ and $(F_\tup{A4},S^\tup{real}_\tup{A4})$, is performed by processing the mainstream data. The values of $F_\tup{A4}$ and $F_\tup{A2}$ are very similar and thus we used the ones in \eqref{eq:F_val}. The attained quantities are noisy, yet they suffice to identify $\delta_{(i,j)}$ and $\beta^\tup{s}_i$ in the model. An in-depth discussion on the model identification is beyond the scope of this paper.   

\subsection{$\tup{A}2$ southbound, from Amsterdam towards Utrecht}

In the A2 stretch, the \gls{ST} is enclosed between two  large junctions, with multiple on- and off-ramps, which brings major perturbations to the flows. 

As can be seen in Table~\ref{tab:a2_cells}, the highway stretch  presents a bottleneck right before the current station location, which means that the infrastructure can offer almost no benefit for the congestion. The $N=15$ cells selected encompass the \gls{ST} that has as access and exit points cells $11$ and $13$, respectively, i.e.,   
$[S^{\tup{real}}_\tup{A2}]_1=11$ and $[S^{\tup{real}}_\tup{A2}]_2=13$.
\begin{table}[t]
\captionsetup{width=\columnwidth}
\begin{center}
\caption{Highway stretch parameters identified for a single lane for the $N=15$ cells in the Dutch $A2$, i.e., $P_{\tup{A2}}$.}
\label{tab:a2_cells}
\begin{tabular}{rccccc}
Cell & ${L}$ & ${\overline v}$ & ${w}$& ${q^{\max}}$ & ${\rho^{\max}}$\\
 & [km] & [km/h] & [km/h] & [veh/h] & [veh/km] \\
\hline
1           & 0,65       & 103        & 31         & 1870           & 79               \\ 
2           & 0.56       & 103        & 25         & 1735           & 86               \\ 
3           & 0.61       & 103        & 33         & 1876           & 75               \\ 
4           & 0.23       & 103        & 26         & 1757           & 84               \\ 
5           & 0.34       & 103        & 33         & 1780           & 71               \\ 
6           & 0.54       & 103        & 35         & 1847           & 71               \\ 
7           & 0.29       & 103        & 38         & 1985           & 72               \\ 
8           & 0.31        & 103         & 40         & 2092           & 73               \\ 
9           & 0.59       & 103         & 40         & 2002           & 69               \\
10          & 0.60        & 96        & 29         & 1714           & 77               \\ 
$[S^{\tup{real}}_\tup{A2}]_1=11$       & 0.41        & 96        & 29         & 1705           & 76               \\ 
12          & 0.20       & 103        & 33         & 1845           & 74               \\ 
$[S^{\tup{real}}_\tup{A2}]_2=13$       & 0.70       & 103        & 35         & 1924           & 74               \\
14          & 0.53       & 104        & 30         & 1774           & 77               \\
15          & 0.51       & 103        & 27         & 1789           & 83               \\ \hline
\end{tabular}
\end{center}
\end{table}
Given $(P_{\tup A2},F_{\tup A2})$ and the flow $\phi_1(k)$, we use both the \gls{GA} and the \gls{NN} --previously trained-- to compute the optimal parameters of the \gls{ST}. The \gls{GA} solves  the problem in approx $15$ min attaining $S^\star_\tup{A2,GA}$, while the \gls{NN}  computes as minimizer of \eqref{eq:NLP}  $S^\star_\tup{A2,NN}$ in less than $0.2$ s. The two vectors of parameters almost coincide, and thus we denote both of them  by $S^\star_\tup{A2}$.
\begin{table}[t]
\captionsetup{width=\columnwidth}
\begin{center}
\caption{Comparison between the features of the real station and those of the computed optimal one on the A2 stretch and comparison between the mixed cases.}
\label{tab:cs_a2}
\begin{tabular}{c|cccc|cc}
& {${i}$} & {${j}$} & {${\delta_{(i,j)}}$} [min] & {${\beta^{\tup s}_i}$} & {${\xi_\Delta}$} [min]  & {${\pi_\Delta}$} \\ \hline
$S^\tup{real}_\tup{A2}$ & 11 & 13 & 80 & 0.10 & \cellcolor{red!20}271.3 & \cellcolor{red!20}0.12 \\ 
$ S^{\square}_\tup{A2}$ & 11 & 13 & 95 & 0.19 & \cellcolor{orange!20}255.3 & \cellcolor{orange!20}0.19 \\ 
$ S^\sqbullet_\tup{A2}$ & 4 & 6 & 80 & 0.10 & \cellcolor{yellow!20}117.5 & \cellcolor{yellow!20}0.26  \\
$S^\star_\tup{A2}$ & 4 & 6 & 95 & 0.19 & \cellcolor{green!20}101.1 & \cellcolor{green!20}0.31  \\
$\bar S^\star_\tup{A2}$ & 7 & 9 & 100 & 0.19 & \cellcolor{yellow!20}183.8 & \cellcolor{orange!20}0.17  \\\hline
\end{tabular}
\end{center}
\end{table}
By inspecting $S^\star_\tup{A2}$ one can notice that the optimal configuration requires $\beta\textsuperscript{s}=0.19$, that is an increment of the $9\%$ with respect to the real value $[S^\tup{real}_\tup{A2}]_4$. This was to be expected, in fact a higher percentage of driver stopping at the \gls{ST} implies a higher damping of the peak traffic congestion. This value, even thought it is high (but lower than the upper bound imposed on $\beta\textsuperscript{s}$), is in line with the data attained from some real \gls{ST}. To achieve the optimal performance in terms of traffic congestion alleviation, the drivers stopping should spend on average more time at the \gls{ST}, specifically $15$ additional minutes.

The major difference highlighted by the proposed algorithm between the optimal \gls{ST} and the real one is the location. The optimal station should be placed around $4$ km before the real one, i.e., between cells $4$ and  $6$, see Table~\ref{tab:cs_a2}. Remarkably, the proposed $S^\star_\tup{GA}$ achieves a reduction of more than $62\%$ of $\xi_\Delta$ and an increment of $19\%$ of $\pi_\Delta$.   
However, the proposed $S^\star_\tup{GA}$ would bring the station in correspondence of a  junction, where the A2 crosses over a smaller road, with  on- and off-ramps. Placing a \gls{ST} in such a location would be unpractical, even though theoretically possible. This highlights a shortcoming of the \gls{NN} with respect to the \gls{GA}. In fact, if a policymaker finds this sort of problem, then he can simply run again the \gls{GA} including this additional constraint. Unfortunately, this cannot be done with the \gls{NN} approach without performing again the training step. The new \gls{GA} solution $\bar S^\star_\tup{A2}$, in which cells $\{4,5,6\}$ cannot be chosen as entry or exit points, still achieves satisfactory results, namely $\xi_\Delta=181$ min and $\pi= 0.17$, see Table~\ref{tab:cs_a2} for comparison. Interestingly, also in this case, the optimal placement of the \gls{ST} lays just before the junction. This shows some level of robustness of the proposed solution, since minor changes in the \gls{ST} position do not abruptly decrease the performance.
%The most obvious element is that the \gls{NN} suggests $\beta\textsuperscript{s}$ to be close to 0.19, which is pretty much at the theoretical limit we found for realistic cases. Having more than 20\% of the mainstream flow entering the station would not only be unrealistic, but even detrimental, as the stopping time of such a large portion of the flow would harm the travel time. The suggested $\delta^{\textup{s}}$ is instead quite interesting: one may think that the highest stopping time would yield the best results, however here the optimal result is just over one and a half hours. What is more surprising is the deep difference in placement. The optimal station is placed around 4 km earlier than the real one, i.e. between cell 4 and cell 6, as shown in table \ref{tab:cs_a2}. 
% This present a big criticality, as it would bring the station in correspondence of a major junction, where the A2 crosses over a smaller road, with an on and off ramp on both sides of the highway. However with some adjustments the station could still yield good results. For example, it could be placed right before or just after the intersection: the simulations ran as such show just about the same $\pi_\Delta$, but an $\xi_\Delta$ that is higher of a couple hundred units. Not a bad result at all, still much better than the current implementation. Moreover, a very common practice is to have on-ramps or off-ramps coincide with ramps to or from a station; this solution would allow to place the station in the exact optimal spot, but would require some revising of the intersection's structure. 

Finally, we complete the case study by analyzing the effect that applying only partially $S^\star_\tup{A2}$ has on traffic congestion.  We denote the case in which  only $\beta^\tup{s}$ and $\delta_{(i,j)}$ are as in $ S^{\star}_\tup{A2}$ by $ S^{\square}_\tup{A2}$. This represents the scenario in which the policymaker is able to change drivers' behaviour at the \gls{ST}, viz. $\beta^\tup{s}$ and $\delta_{(i,j)}$, for example via incentives, but not the \gls{ST} location. With respect to $S^\tup{real}_{A2}$, we obtain a smaller reduction of $\xi_\Delta$ while the effect in terms of peak traffic reduction remains sensible, since there is an increment of $7\%$ of $\pi_\Delta$. In the second case, the position of the \gls{ST} can be chosen as $S^{\star}_\tup{A2}$ but the drivers' behavior is not affected, we denote it by $ S^\sqbullet_\tup{A2}$. The performance reduction with respect to $S^{\star}_\tup{A2}$ is minor, see Table~\ref{tab:cs_a2}, this highlights that the location of the \gls{ST} has a bigger effect on traffic congestion compared to how many drivers stop and for how long.

% The mixed cases are possibly more interesting than the optimal one, indeed the impact of varying only $\beta\textsuperscript{s}$ and $\delta\textsuperscript{s}$, while leaving the station where it is in the real world is limited, but not null, meaning some gains are possible with an active control on $\beta\textsuperscript{s}$ and $\delta\textsuperscript{s}$. On the other hand, when the placement is optimal, the effect of modifying $\beta\textsuperscript{s}$ and $\delta\textsuperscript{s}$ is much greater, so the reasonable conclusion is that both optimizations must be applied at the same time for the maximum benefit.

\subsection{$\tup{A}4$ eastbound, from Amsterdam towards Den Haag}

A4 highway stretch has been selected for the interesting configuration of its ramps, indeed the on-ramp before the station merges into the mainstream with a long accessory lane, which then seamlessly turns into the lane for accessing the \gls{ST}. The same happens for the lane exiting the \gls{ST}, becoming an off-ramp after the station. % This unusual disposition makes for an interesting case to study, as this setup may lead to unexpected challenges. 
Also in this case, the stretch has been divided into $N=15$ cells identified by $P_\tup{A4}$, where the \gls{ST} entering point  is at cell $7$ and exiting point at cell $9$, see Table~\ref{tab:a4_cells}. 

\begin{table}[h]
\captionsetup{width=\columnwidth}
\begin{center}
\caption{Highway stretch parameters identified for a single lane for the $N=15$ cells in the $A4$, i.e., $P_{\tup{A4}}$.}
\label{tab:a4_cells}
\begin{tabular}{rccccc}
Cell & ${L}$ & ${\overline v}$ & ${w}$& ${q^{\max}}$ & ${\rho^{\max}}$\\
 & [km] & [km/h] & [km/h] & [veh/h] & [veh/km] \\
\hline
1           & 0,31       & 113        & 21         & 1730           & 96               \\ 
2           & 0.38       & 114        & 27         & 1765           & 80               \\ 
3           & 0.56       & 114        & 28         & 1794           & 79               \\ 
4           & 0.49       & 114        & 25         & 1793           & 87               \\ 
5           & 0.31       & 113        & 32         & 1989           & 79               \\ 
6           & 0.41       & 112        & 32         & 2005           & 80               \\ 
$[S^{\tup{real}}_\tup{A4}]_1=7$           & 0.44       & 112        & 56         & 2148           & 58               \\ 
8           & 0.42        & 111         & 55         & 2513           & 68               \\ 
$[S^{\tup{real}}_\tup{A4}]_2=9$           & 0.33       & 109         & 55         & 2296           & 63               \\ 
10          & 0.56        & 108        & 37         & 2011           & 72               \\ 
11          & 0.34        & 108        & 35         & 2009           & 74               \\ 
12          & 0.26       & 109        & 38         & 2030           & 76               \\ 
13          & 0.32       & 108        & 39         & 1999           & 72               \\ 
14          & 0.22       & 108        & 50         & 1999           & 70               \\ 
15          & 0.59       & 108        & 40         & 2000           & 77               \\ \hline

\end{tabular}
\end{center}
\end{table}

\begin{table}[h]
\captionsetup{width=\columnwidth}
\begin{center}
\caption{Comparison between the features of the real station and those of the computed optimal one on the A4 stretch.}
\label{tab:cs_a4}
\begin{tabular}{c|cccc|cc}
& {${i}$} & {${j}$} & {${\delta_{(i,j)}}$} [min] & {${\beta^{\tup s}_i}$} & {${\xi_\Delta}$} [min]  & {${\pi_\Delta}$} \\ \hline
$S^\tup{real}_\tup{A4}$ & 7 & 9 & 65 & 0.11 & \cellcolor{red!20}1304 & \cellcolor{orange!20}0.24 \\ 
$ S^{\square}_\tup{A4}$ & 7 & 9 & 86 & 0.19 & \cellcolor{yellow!20}981 & \cellcolor{yellow!20}0.29 \\ 
$ S^\sqbullet_\tup{A4}$ & 13 & 15 & 65 & 0.11 & \cellcolor{orange!20}1286 & \cellcolor{red!20}0.20  \\ 
$S^\star_\tup{A4}$ & 13 & 15 & 86 & 0.19 & \cellcolor{green!20}806 & \cellcolor{green!20}0.36 \\ \hline
\end{tabular}
\end{center}
\end{table}
Once again, the optimal set of parameters describing the \gls{ST} $S^\star_\tup{A4}$ is computed via the \gls{GA} and the \gls{NN}, and  they {are almost} identical.
In this case, the optimal \gls{ST} position lays just after the off-ramp mentioned above, i.e., $[S^\star_\tup{A4}]_1=13$ and {$[S^\star_\tup{A4}]_2=15$}. The ideal split ratio $[S^\star_\tup{A4}]_4=0.19$ coincides with the one identified in the previous case study, confirming the intuition  that a higher percentage of vehicles stopping at the \gls{ST} creates a beneficial effect in terms of traffic mitigation. Moreover, to achieve the optimal performance, the time spent at the \gls{ST} should also increase of $20$ min. 
The improvements are milder than in the case of the A2, but still relevant, both in terms of total congestion time saved, and peak of traffic congestion reduction. In fact, $\xi_\Delta$ is reduced of $39\%$ and $\pi_\Delta$ grows of $12\%$, see Table~\ref{tab:cs_a4}. 

% \red{Can we redo the sims shifting the 15 cells forward of a bit? Cause otherwise it is on the edge that seems a bit sketchy as result.}
% In this case, the suggested configuration lands the station just after the off-ramp mentioned above. This is not entirely surprising, as on and off-ramps are sources of perturbation on the road, so having a \gls{ST} between these two points is probably little to no good when it comes to congestion reduction. Once again we can notice how the ideal split ratio is the same as before, close to the theoretical maximum, and the stop time is similar as well. The gains in this case are slightly more marginal, but still very significant, both in terms of total congestion time saved and of profile of the congestion. In this case, the proposed positioning is slightly less problematic than the first situation studied. As already mentioned, the input ramp for the station could be made to coincide with the off-ramp of the highway, and the output ramp would just cross over or under the exit road that is already in place. Also, no particular feature of the landscape would suggest any difficulty in building an infrastructure here, other than obviously obtaining the permits and concessions to use the land. 
As for the A2 case, we consider $ S^\sqbullet_\tup{A4}$ and $ S^\square_\tup{A4}$  a partial implementation of the optimal \gls{ST} design in $ S^\star_\tup{A4}$. The case in which only the position of the service station changes, i.e., $ S^\sqbullet_\tup{A4}$,  is of particular  interest. In fact, the total congestion time $\xi_\Delta$ decreases but the maximum peak of congestion increases, i.e., $\pi_\Delta$ decreases, with respect to the current situation. This highlights that the choice  of the position of the \gls{ST} and the policies influencing  drivers' behavior  are not decoupled. On the contrary, they work in synergy. Therefore,  if one is interested in finding only the optimal \gls{ST} position, then a different \gls{GA} should be implemented in which $(\beta^\tup{s},\delta_{(i,j)})$ are moved from the vector of the optimization variables $S$ to the set of fixed parameters $F$. The possibility of solving such a complex problem by minor adjustment of the algorithm confirms the power of the proposed optimization scheme.

\section{Conclusion}
The presence and the features of a \gls{ST} on a highway stretch can highly affect the level of traffic congestion throughout the day. The use of microsimulators  to compute a priori its optimal design is not a viable option due to their computational complexity.
We develop a \gls{GA} and a \gls{NN}, based on the macroscopic traffic model \gls{CTMs}. They successfully  calculate the optimal design of the \gls{ST} that minimizes the traffic congestion. We show via two case studies, based on the real data retrieved form the  Dutch highway, the benefit of an {optimally designed} \gls{ST}. Namely, it can halve the level of traffic congestion if the \gls{ST} is placed correctly and enough drivers stop for enough time. These algorithms give valuable insights to policymakers that can be later refined with more demanding simulation tools. They also provide an indication of which policies should be put in place to make the \gls{ST} operate in optimal conditions. 

The \gls{CTMs} can be easily extended to the multi-modal case, where different types of vehicles travel  and  use  the \gls{ST} differently. It is interesting to extend our approach to this more complex scenario. Moreover, we are interested in extending the simulations to a higher number of cells and increase the number of \glspl{ST} to be designed. We think that our algorithms can produce interesting insights. %  in this scenario.
\bibliography{ifacconf}             % bib file to produce the bibliography
                                                     % with bibtex (preferred)

% \appendix
% \section{A summary of Latin grammar}    % Each appendix must have a short title.
% \section{Some Latin vocabulary}              % Sections and subsections are supported  
%                                                                          % in the appendices.
\end{document}